\documentclass[a4paper]{article}

\pdfoutput=1

\usepackage{amsmath}
\usepackage{amssymb}
\usepackage[utf8]{inputenc}
\usepackage{booktabs}
\usepackage{url}
\usepackage[symbol]{footmisc}
\usepackage{perpage}
\usepackage{algorithm}
\usepackage{algpseudocode}
\usepackage{multirow}
\usepackage{graphicx}
\usepackage{caption}
\usepackage{subcaption}
\usepackage{wrapfig}
\usepackage{dsfont}
\usepackage{a4wide}
\usepackage{authblk}

\usepackage{ifthen}	
\newboolean{showcomments}
\setboolean{showcomments}{true}
\ifthenelse{\boolean{showcomments}}
{ \newcommand{\mynote}[2]{
    \fbox{\bfseries\sffamily\scriptsize#1}
    {\small$\blacktriangleright$\textsf{\emph{#2}}$\blacktriangleleft$}}}
{ \newcommand{\mynote}[2]{}}

\newcommand{\lib}[1]{\textit{\lowercase{#1}}}

\newcommand{\figref}[1]{Figure~\ref{fig:#1}}
\newcommand{\tabref}[1]{Table~\ref{tab:#1}}
\newcommand{\secref}[1]{Section~\ref{sec:#1}}

\title{\LARGE \bf A Study of Library Migration in Java Software}
\date{}

\author{Cédric Teyton}
\author{Jean-Rémy Falleri}
\author{Marc Palyart}
\author{Xavier Blanc}
\affil{Univ. Bordeaux, LaBRI, UMR 5800, F-33400 Talence, France\\\{cteyton,falleri,mpalyart,xblanc\}@labri.fr}

\begin{document}

\maketitle

\begin{abstract}
Software intensively depends on external libraries whose relevance may change during its life cycle. As a consequence, software developers must periodically reconsider the libraries they depend on, and must think about \textit{library migration}. To our knowledge, no existing study has been done to understand library migration although it is known to be an expensive maintenance task. Are library migrations frequent? For which software are they performed and when? For which libraries? For what reasons? The purpose of this paper is to answer these questions with the intent to help software developers that have to replace their libraries. To that extent, we have performed a statistical analysis of a large set of open source software to mine their library migration. To perform this analysis we have defined an approach that identifies library migrations in a pseudo-automatic fashion by analyzing the source code of the software. We have implemented this approach for the Java programming language and applied it on Java Open Source Software stored in large hosting services. The main result of our study is that library migration is not a frequent practice but depends a lot on the nature of the software as well as the nature of the libraries. 
\end{abstract}

\section{Introduction}
\label{sec:intro}

Almost all software applications depend on external libraries that provide useful technical facilities. Examples of such libraries are \lib{JUnit} for unit testing or \lib{Log4J} for logging. The relevance of a library for a software project may change during its life cycle. As a consequence, software developers must periodically reconsider the libraries they depend on, and must think about library migration when the libraries they depend on are not updated, or when competing ones appear with more features or better performance for instance. 

To the best of our knowledge, no existing study has been done to understand library migration although it is known to be a highly costly maintenance task. Are library migrations frequent? For which kinds of software are they performed and when? For which libraries? For what reasons? etc. The purpose of this paper is then to answer these questions with the intent to help any software developer that will have to think about replacing the libraries she uses.

In particular, we propose to address the following research questions:
\begin{enumerate}
\item Which software projects perform library migrations and how much of them? Our objective is to check if some kinds of software are prone to library migrations.
\item Which libraries are migrated and how many times? Our objective is to check if some libraries are prone to be source or target of migrations.
\item When migrations are performed? Our objective is to check if there is any tendency regarding migrations. 
\item Why migrations are performed? Our objective is to identify the common causes of library migrations.
\item How migrations are performed? Our objective is to measure the mean time that is needed to perform a migration with the intent to serve as an indication for assessing migration effort costs.
\end{enumerate}

Answering these research questions would help the software developers thinking about migrating their libraries. Currently, for such intent they can only use general purpose search engines, such as Google, which give only partial and sometimes out-of-date answers. For example, if a developer wants to migrate its \lib{Commons-logging} library, she will probably query Google with something similar to: ``\textit{logging library Java}''. She will obtain a list of technical websites but no advice that would help her find an adequate replacement library neither any pointer to existing software that did already perform such a migration.

In this paper, we answer these research questions by analyzing a large corpus of Java software. Our objective is to mine a large set of library migrations to statistically exhibit common practices. For instance, we assume that if we observe that a large number of software project migrate from \lib{Log4J} to \lib{SLF4J}, it is then relevant for every similar software project using \lib{Log4J} to consider \lib{SLF4J} as a good candidate to migrate to. 

Our contribution is twofold. First, we propose a mining approach to pseudo automatically identify library migrations that occur in Java software. Second, we answer our research questions with results from a study performed with our mining approach on three major hosting platforms (GitHub, Google Code and Source Forge). 

This paper extends our previous work~\cite{teyton_mining_2012} which only targets software that use the build automation tool Maven. The study we present in this paper takes into account any kind of Java Open Source Software. The results we obtain here are therefore more general than the ones from our previous paper.

The remainder of this paper is structured as follows. \secref{identifying_migrations} first explains our approach to identify library migrations and \secref{study} presents the study we performed on projects stored on three major hosting platforms. \secref{research_questions} then presents the answers to the research questions. \secref{limits} discusses the limitations of our approach. \secref{rw} presents the related work, while \secref{conclusion} uncovers the future work and concludes.

\section{Identifying migrations}
\label{sec:identifying_migrations}

\newtheorem{thm}{Theorem}
\newtheorem{definition}{Definition}

In this section, we first introduce the abstract model we define to represent software projects and library migrations. Based on this model, we present the approach we use to identify library migrations. 

\subsection{Dependency Model}
\label{sec:identifying_migrations:dependency_model}

We abstract the data needed to perform our analysis in a dependency model. This model is very simple as it only contains the set of analyzed software projects, their list of versions and, for each version the associated set of library dependencies. 

\begin{definition}[Software projects and Libraries]
Let $P$ be the set of software projects and $L$ be the set of libraries. 
For the sake of simplicity, we consider that each project $p \in P$ has an associated totally ordered set $V_p \subset \mathds{N}$ of versions. Versions are sorted chronologically according to their date. 
For a project $p \in P$ at version $i \in V_p$, we define $dep_p(i) : V_p \rightarrow {\cal P}(L)$ the set of its library dependencies.
\label{def:projects-libraries}
\end{definition}

Let us illustrate our model with an example. We assume two projects ($\mathtt{P_A}$ and $\mathtt{P_B}$) and four libraries (\lib{JUnit}, \lib{TestNG}, \lib{Log4J} and \lib{SLF4J}). Table \ref{tab:dependency-model} presents this dependency model with versions of $\mathtt{P_A} : (1, 2, 3)$ and of $\mathtt{P_B} : (1, 2)$, associated with their corresponding dependencies. Note that version $1$ of $\mathtt{P_A}$ and version $1$ of $\mathtt{P_B}$ are different and occur at two different dates (as well as for the versions $2$). 

\begin{table}[h]
\centering
\scriptsize
\begin{tabular}{cccc}
\toprule
Project & \multicolumn{3}{c}{Versions / Dependencies}\\
\midrule
$\mathtt{P_A}$ & $1$ & $2$ & $3$\\
& $\{$\texttt{JUnit}$\}$ & $\{$\texttt{JUnit},\texttt{TestNG},\texttt{Log4J}$\}$& $\{$\texttt{TestNG},\texttt{SLF4J}$\}$\\
\midrule
$\mathtt{P_B}$ & $1$ & $2$ & \\
& $\{$\texttt{TestNG}$\}$ & $\{$\texttt{JUnit}$\}$&\\
\bottomrule
\end{tabular}
\caption{Example of projects with their versions and their dependencies}
\label{tab:dependency-model}
\end{table}

\begin{definition}[Library migration]
We state that a project $p \in P$ migrates from a library $s \in L$ to another library $t \in L$ if it depends on $s$ at version $i \in V_p$ (i.e. $s \in dep_p(i)$) and if this dependency is replaced by $t \in L$ in version $j \in V_p$ with $i < j$ (i.e. $t \in dep_p(j)$). A migration is therefore a tuple $(p,i,j,s,t) \in P \times \mathds{N} \times \mathds{N} \times L \times L$. Finally, given a set of projects $P$ and a set of libraries $L$, we note $M$ all the migrations that occur for all projects in $P$.
\label{def:library-migration}
\end{definition}

Regarding our sample presented in Table \ref{tab:dependency-model}, the following migrations have been performed in projects $P_A$ and $P_B$: $(P_A,1,3,junit,testng)$, $(P_A,2,3,log4j,slf4j)$ and $(P_B,1,2,testng,junit)$. 

\begin{definition}[Library migration rule]
We also call a migration rule a couple $(s,t) \in L^2$ such that there exists at least one project $p \in P$ that migrates from $s$ to $t$ during its life cycle. We note $R$ the set of all library migration rules.
\label{def:library-migration-rule}
\end{definition}

Regarding our example $R=\{(junit,testng), (log4j , slf4j), (testng , junit)\}$.

\subsection{Mining library migrations}
\label{sec:identifying_migrations:mining_lib_migrations}

To answer our research questions, we have to identify migrations $M$ that occur in a set of projects $P$ for a set of libraries $L$. Such an identification obviously first requires to define $P$ but also $L$. While defining $P$ is straightforward, defining $L$ is not so obvious and mainly depends on the programming language. \secref{study} presents how we manage to identify $P$ and $L$ for Java Open Source projects.

Once $P$ and $L$ have been defined, our approach has to describe how library dependencies can be automatically computed (the $dep_p(i)$ function has to be defined). To that extent, several techniques, such as the ones that are based on tools that manage library dependencies like Maven ~\cite{teyton_mining_2012}, can be used. In this paper, we choose not to depend on a specific tool but rather to use source code static analysis for automatically computing actual dependencies between a project and a set of libraries. \secref{study} presents such a simple static analysis for Java projects and Java libraries.

Further, to compute $M$, we propose an algorithm that iterates on several versions of each project to identify migrations that may have been performed. In detail, our algorithm looks at couples of versions $(i,j)$ in a project $p$ and checks which library dependencies existed at version $i$ but were replaced at version $j$. Our algorithm returns a candidate migration for each element of the Cartesian product between the dependencies that existed at $i$ but were removed at $j$ and the ones that did not exist at $i$ but were added at $j$. 

With our example, considering the project $P_A$ and the couple of versions $(1,3)$, our algorithm returns the candidates $(P_A,1,3,junit,testng)$ and $(P_A,1,3,junit,slf4j)$. Considering the same project but the couple of versions $(1,2)$, our algorithm does not return any candidate. Finally, with the same project but the couple of versions $(2,3)$, the candidates $(P_A,2,3,junit,testng)$, $(P_A,2,3,junit,slf4j)$, $(P_A,2,3,log4j,testng)$ and $(P_A,2,3,log4j,slf4j)$ are returned. 

\begin{figure}[h]
		\centering
        \includegraphics[width=0.7\linewidth]{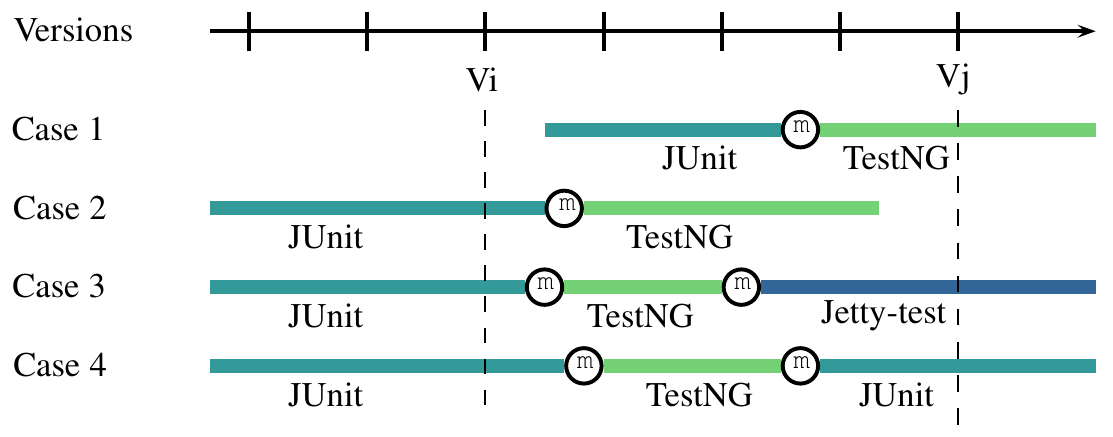}
        \caption{Cases of migrations missed if the distance between the versions couple is too large}
		\label{fig:largeDistanceCouple}
	\end{figure}

The choice of couples $(i,j)$ to observe has a major impact on the quantity of the candidate migrations returned by our algorithm. Ideally all couples should be observed to get all possible candidates but this takes too much time and is certainly not useful as many couples return the same candidate migrations. Further, the choice of the distance between the two versions of the couple ($j-i$) has also an impact on the returned candidates. The following two cases should be considered: 
	\begin{itemize}
		\item \textit{Large} distance couples ($(j-i)$ approaches to $\infty$). If the distance between the couple is too large, several migrations that occur between $i$ and $j$ will not be considered as candidates. The \figref{largeDistanceCouple} presents four situations in which such a case occurs. In this figure, a horizontal axis represents a dependency toward a certain kind of library for a given project $p$ (for instance a dependency toward a testing library). Each colored segment displays which library has been used as a dependency (e.g. $junit$ then $testng$) and thus also displays when migrations occurred (\textcircled{m}). Each of the four cases represents distinct situations which are:  
			\begin{enumerate}
				\item The introduction of $junit$ was done after $i$, hence by observing the couple $(i,j)$ our algorithm does not know that $junit$ was used before $testng$ and therefore will not return any candidate.
				\item The project stopped to use $testng$  before $j$, thus by observing the couple $(i,j)$ our algorithm does not know $testng$ and therefore will not return any candidate. 
				\item Two migrations happened between $i$ and $j$. $junit$ was replaced by $testng$ that was then replaced by $jetty-test$. Here, by observing the couple $(i,j)$ our algorithm will consider only the candidate $(p,i,j,junit,jetty-test)$ even though this migration never happened directly. As a matter of fact, if the distance is too large our algorithm might return migrations that are the result of combinations of several successive real migrations. 
				\item The project migrated from $junit$ to $testng$ between $i$ and $j$, but then moved back to $junit$ before the version $j$. By observing the couple $(i,j)$ our algorithm will not return any candidate.
			\end{enumerate}	 
		\item \textit{Small} distance couples ($(j-i)$ approaches to $1$). If the distance between the couple is too small, migrations that are performed during several versions cannot be detected. Indeed, we have no clue that a migration is always performed during only one version. The new dependency might be added at version $i$ but the old one might be kept at $j$ before being removed at $j+k$. Our example highlights such a situation with the project $S_A$. In particular, if our algorithm observes the couple of versions $(1,2)$ and $(2,3)$, the candidate $(S_A,1,3,junit,testng)$ will not be returned.
	\end{itemize}
	
The choice of the couples to observe and their distance therefore requires to reach a trade-off between the quantity of returned migrations and time efficiency. \secref{study} presents a discussion about this trade-off for Open Source Java projects.

Whatever the number of observed couples of versions and whatever their distance, once candidate migrations have been identified, they have to be validated either automatically or manually before to be truly included in $M$, the set of migrations. Regarding our example and the couple $(1,3)$, the candidate $(P_A,1,3,junit,slf4j)$ should be filtered out while $(P_A,1,3,junit,testng)$ should be filtered in. \secref{study} explains that, albeit an automatic validation provided good results in our previous study, the recall suffered from the priority given to precision. We decide in this study to focus on recall only and this is why we have realized a manual validation to discard the irrelevant candidates.

\section{Experiments}
\label{sec:study}

Based on the approach described in the previous section we conducted a case study on Java Open Source software. We have focused our study on the Java programming language for the sake of simplicity but it can be easily extended to any other programming language. 

This section presents how we operated to apply our approach for this study. First, we introduce how we built $P$ the corpus of software projects and $L$ the set of libraries. Second, we describe how for a given project version we are able to identify libraries it uses. Finally, we explain how the migration rules are detected. 

\subsection{The set of projects $P$}
\label{sec:study:P}

A corpus of 14,028 software projects has been built by querying Github, GoogleCode and Sourceforge hosting platforms to get the Java projects they manage. The selection was then achieved in a random manner. Among these projects we first removed 2,430 empty repositories that did not contain Java source code or were merely not existing anymore. Then, we discarded 2,380 projects that did not use any third-party library. We finally obtained a set $P$ composed of 9,218 projects. 

\subsection{The set of libraries $L$}
\label{sec:study:L}

A set of Java libraries was reused from a prior study \cite{teyton_mining_2012}. In this previous experiment we analyzed over 6,000 Java projects managed with Maven. A Maven command was used to obtain libraries JAR files used by each project. Thanks to this process we gathered a base of 8,795 libraries in the form of JAR files. We then grouped these files according to the similarity of their names. For instance, we consider that \textit{junit-4.8.1.jar} and \textit{junit-4.8.2.jar} are part of the same group, that corresponds to the \lib{JUnit} library. After this operation we obtained 3,326 libraries. 

To detect library dependencies using source code static analysis, we consider that a library is identifiable by the packages it defines. Thus, for each library we built a set of regular expressions that matches its package names. The construction of these sets has been done by analyzing the 3326 JAR files with the bytecode engineering library \textsc{Javassist} \cite{chiba_easy--use_2003}. We however faced many issues while performing this operation. First, we observed that even if some JAR files have different names and seems to belong to different libraries, they belong in reality to the same library. For example, the library \lib{batik} from Apache is composed of \lib{batik-svggen}, \lib{batik-dom}, \lib{batik-script} and other components. That is why in such case we decided to manually group them. Second, we also observed that different libraries might define packages with similar names. For instance, the package name \textit{"com.google.common.io"} is found in both \lib{guava} and \lib{craftbukkit} libraries. When such case occurs, a manual intervention is required to state in which library the package has to be assigned or if it has to be not included at all. To fix all these issues, one person spent about 14 hours to manually review the computed regular expressions. After this operation we obtained a set $L$ of 1189 libraries with a set of regular expressions for each one of them. The current index of $L$ is available on-line\footnote{\url{http://www.labri.fr/perso/cteyton/ScanLib/scanlib.html}}.

\subsection{Detecting library dependencies}
\label{sec:study:depend}

Thanks to the regular expressions built throughout the construction of the set $L$, the detection of library dependencies for a given project version is straightforward. A static analysis attempts to match the regular expressions of libraries with the import sections or the qualified names used within the source code of the Java files. This analysis has the advantage to be very simple and requires only a few seconds to compute the library dependencies of a project. The Eclipse JDT parser is used to traverse the AST of Java source code files. All this process is implemented by our tool, named \textsc{ScanLib}, which is available on-line\footnote{\url{https://code.google.com/p/scanlib-java/}}. This tool is written in Java and runs along the database of libraries $L$ described above.

\subsection{Detecting migrations}
\label{sec:study:M}

\begin{figure}[h]
	\centering
	\includegraphics[width=0.85\linewidth]{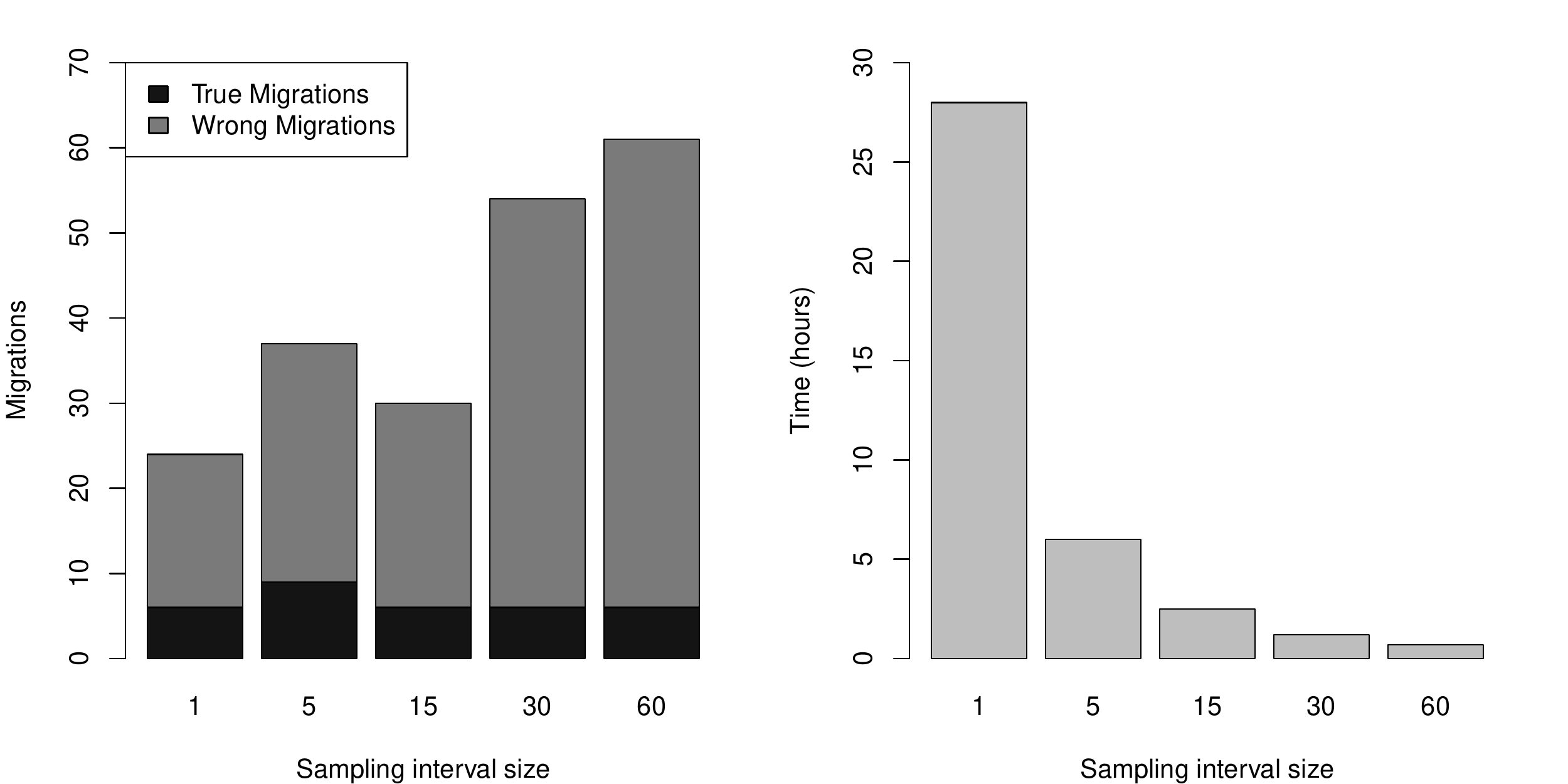}
	\caption{Results of the experiments for different sampling interval sizes}
	\label{fig:sampling_results}
\end{figure}

As we described it in the Section \ref{sec:identifying_migrations:mining_lib_migrations}, our algorithm selects a set of couples of versions to identify candidate migrations. In this study we decided to choose sequential versions as couples with a fixed step as a distance between the versions. For example with a step of 10, the couples are $\{(1,11),(11,21),(21,31),..., (n,n+10)\}$. To determine such step, we selected randomly 150 projects from our original corpus and compared the candidates returned with different steps: 1, 5, 15, 30 and 60. For each step, we measured the time taken to produce the complete analysis, the number of returned candidate migrations, and how many true and false positives were generated among the candidates. The resulting data are exposed in \figref{sampling_results}. On this sample of projects, the best sensibility is obtained with a step of 5 versions.  However, we decided not to choose this step because first our experiment does not guarantee that this value is also the optimum for our global corpus, and second by making a projection of the execution time we find that several weeks of computation would be necessary. That is why we have considered that an interval of 30 versions was a good trade-off between the number of migrations we would gather and an execution time below one week. 

Once we selected 30 as our step, we looked at each project in the corpus $P$. For a given project we first checked out its repository to its first version, and then synchronized the local repository every 30 versions as long as a new step was reachable. In the opposite case, the last commit version is considered. At each step, the list of qualified names is fetched using the static analysis tool included in \textsc{ScanLib}. The output collection of names is stored, along a timestamp information in order to date the events. To implement this process we developed a prototype based on the Harmony framework \footnote{\url{https://code.google.com/p/harmony/}}. Harmony is an infrastructure designed to ease the development of tools for software evolution analysis. The execution of this process with the prototype required about one full week to complete and produced 4,9 GB of data.

A post-processing operation executed \textsc{ScanLib} on the data to resolve whether the qualified names extracted from the Java source code could match any library from $L$. This task took about 8 minutes. At the end of this step we were able to compute the $dep_p()$ function on any analyzed version of any project. Then we applied the Cartesian product on the added and removed dependencies of every couple of versions from projects in $P$. This operation produced in 20 seconds a set $M$ of 3,579 migrations that can be grouped according to 2,920 migrations rules (set $R$).

Obviously the set $R$ contains a large quantity of false positive rules. As detailed earlier, the data mining process proposed in \cite{teyton_mining_2012} to overcome false positives is not renewed here. In our opinion, it is worthwhile to dedicate efforts to manually rate the migrations rules in order to keep all the true migrations rules. One person spent 4 hours to manually check the candidate migration rules and rated them either as correct or wrong. Naturally, when a migration rule from the set $R$ was marked as wrong, all the migrations following this rule were removed from the set $M$. At the end of the validation process we obtained a set $M$ of 342 migrations that can be grouped according to 134 migrations rules (set $R$). The full list of migration rules is available online \footnote{\url{http://www.labri.fr/perso/cteyton/index.php?page_name=migrations_rules}}.

\section{Research Questions}
\label{sec:research_questions}

This section answers the research questions presented in Section \ref{sec:intro}. For each question, we present the methodology we used to answer the question and then we present the results obtained using the data extracted in \secref{study}.

\subsection{Which software projects perform library migrations and how much of them?}
\paragraph{Methodology.}
To identify the kinds of software that are prone to migration, we decided to use metrics to statistically measure the size of projects and to compare it with the number of performed migrations.

We arbitrary chose three metrics that are classically used to measure the size of a project: KLOC, Commit and Duration. The KLOC metric computes the number of lines of Java code. The Commit metric returns the number of commits that were performed during the life of the project. The Duration metric measures the lifetime of the project in months. 

Once we have measured these three metrics for all the projects, we decided to remove all projects that have less than 100 KLOC, then the ones that have less than 10 commits and then the ones that are younger than one month. Our intent was to remove all toy projects that were not significant regarding library migrations.

Finally, we decided to compute deciles for each metric in order to define groups of projects with similar size. Next we have measured the number of migrations for each group and we have performed a Chi-squared test to check if the distribution of migrations is the same for all groups for each metric.

If the distribution is not the same, we plot the distributions of the groups for each metric to observe how the number of migrations varies depending on the metric.

\paragraph{Results.}

By removing toy projects, we finally obtained 8,600 projects from the 9,218 ones of our experiment. As we observed $342$ library migrations, this means that only $3,9\%$ of software projects perform at least one migration during its life cycle. It therefore appears that library migrations do occur but are very rare. 

We then have computed the deciles for the three metrics and the number of migrations. \tabref{group_kloc}, \tabref{group_commit} and \tabref{group_duration} present the groups for each metric and present how many projects are contained in each group and how many migrations have been performed.

\begin{table}[h]
\centering
\footnotesize{
\begin{tabular}{lcccccccccc}
\toprule
Group & 1 & 2 & 3 & 4 & 5 & 6 & 7 & 8 & 9 & 10 \\
\midrule
\textbf{Java KLOC} &  & $>$ 0.54 & $>$ 0.96 & $>$  1.5 & $>$  2.2 & $>$  3.2 & $>$  4.9 & $>$  7.7 & $>$  13.7 & $>$  30.4  \\ 
 & $\leq$ 0.54 & $\leq$ 0.96 & $\leq$ 1.5 & $\leq$ 2.2 & $\leq$ 3.2 & $\leq$ 4.9 & $\leq$ 7.7 & $\leq$ 13.7 & $\leq$ 30.4 & $\leq$ 3675\\
\midrule
\textbf{Projects} & 851 & 850 & 843 & 838 & 831 & 837 & 838 &  814 & 822 & 812 \\
\midrule
\textbf{Migrations} & 6 & 13 & 17 & 22 & 29 & 22 & 23 & 46 & 38 & 47 \\
\bottomrule
\end{tabular}
}
\caption{The groups for the KLOC metric}
\label{tab:group_kloc}
\end{table}

\begin{table}[h]
\centering
\footnotesize{
\begin{tabular}{lcccccccccc}
\toprule
Group & 1 & 2 & 3 & 4 & 5 & 6 & 7 & 8 & 9 & 10 \\
\midrule
\textbf{Commit} &  & $>$ 25 & $>$ 31 & $>$ 38 & $>$ 49 & $>$ 65 & $>$ 87 & $>$ 127 &  $>$ 207 & $>$ 430 \\ 
 & $\leq$ 25 & $\leq$ 31 & $\leq$ 38 & $\leq$ 49 & $\leq$ 65 & $\leq$ 87 & $\leq$ 127 & $\leq$ 207 & $\leq$ 430 & $\leq$ 29656 \\
\midrule
\textbf{Projects} & 840 & 869 & 791 & 857 & 900 & 829 & 833 & 825 & 815 & 777 \\
\midrule
\textbf{Migrations} & 0 & 3 & 7 & 12 & 18 & 22 & 29 & 40 & 46 & 86 \\
\bottomrule
\end{tabular}
}
\caption{The groups for the Commit metric}
\label{tab:group_commit}
\end{table}

\begin{table}[h]
\centering
\footnotesize{
\begin{tabular}{lcccccccccc}
\toprule
Group & 1 & 2 & 3 & 4 & 5 & 6 & 7 & 8 & 9 & 10 \\
\midrule
\textbf{Duration} &  & $>$ 3 & $>$ 5 & $>$ 6 & $>$ 8 & $>$ 11 & $>$ 15 & $>$ 21 & $>$ 29 & $>$ 45  \\ 
 & $\leq$ 3 & $\leq$ 5 & $\leq$ 6 & $\leq$ 8 & $\leq$ 11 & $\leq$ 15 & $\leq$ 21 & $\leq$ 29 & $\leq$ 45 & $\leq$ 512 \\
\midrule
\textbf{Projects} & 704 & 749 & 769 & 918 & 872 & 892 & 949 &  839 & 838 & 806 \\
\midrule
\textbf{Migrations} & 4 & 4 & 5 & 10 & 17 & 33 & 27 & 52 & 45 & 66 \\
\bottomrule
\end{tabular}
}
\caption{The groups for the Duration metric}
\label{tab:group_duration}
\end{table}

To evaluate if the number of migrations was independent of the group, we used statistical Chi-squared test. The null hypothesis is that the proportion of migrations is the same whatever the group for each metric. This test was therefore computed for the three metrics. Chi-squared test results in \tabref{deciles} suggest that our corpus is significant and that the null hypothesis is rejected with a probability of $5\%$. This means that the proportion of migrations is not the same depending of the group for each metric.

\begin{table}[h]
\centering
\begin{tabular}{lccc}
\toprule
& CLOC & Commit & Duration \\
\midrule
X-square & 66.1 & 235.0 & 158.9 \\
degree of freedom & 9 & 9 & 9 \\
p-value & $8.8*10^{-11}$ & $2.2*10^{-16}$ & $2.2*10^{-16}$ \\
\bottomrule
\end{tabular}
\caption{Results for the Chi-squared test}
\label{tab:deciles}
\end{table}

\begin{figure}[h]
\centering
\includegraphics[width=1.0\linewidth]{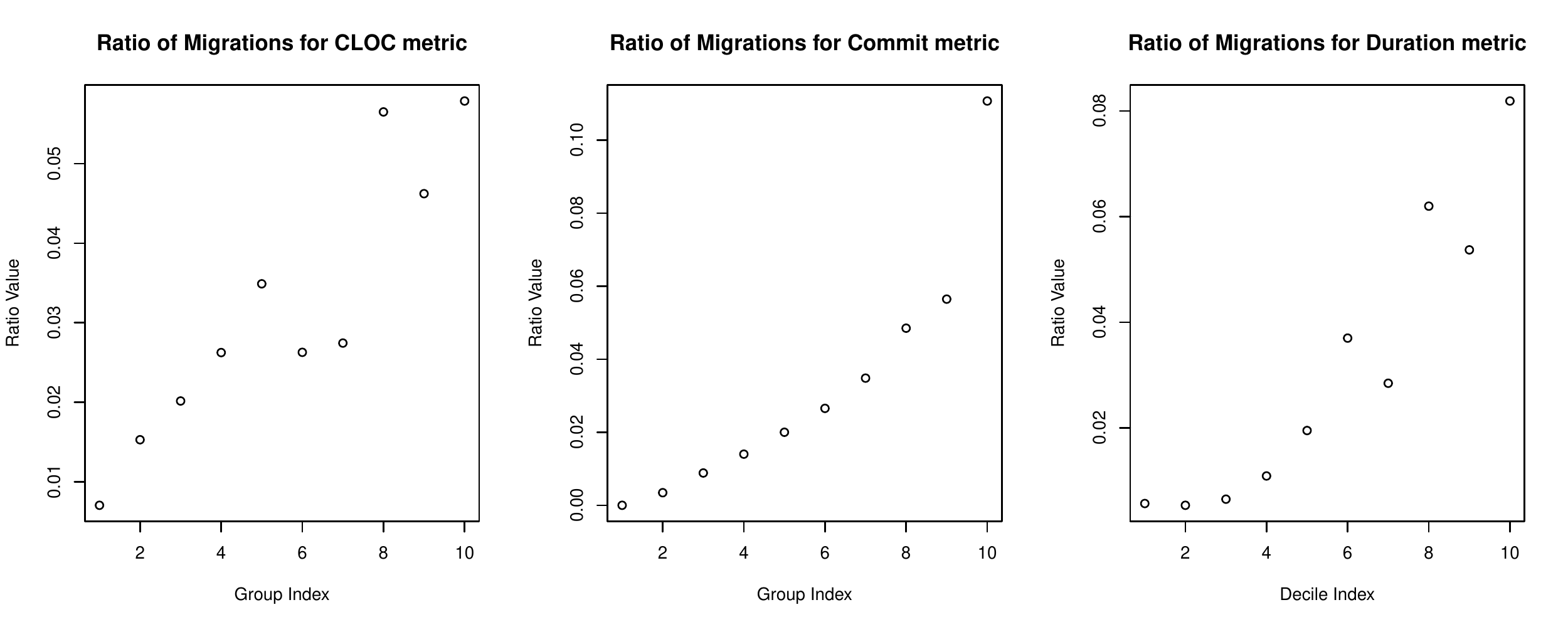}
\caption{Distribution of migrant projects per CLOC, Commit and Duration deciles}
\label{fig:ratio}
\end{figure}

To visualize the differences in proportion of each group, we plot in \figref{ratio} the ratio for each metric (number of migration per number of projects). This Figure clearly shows that the proportion of migrations is more important in the biggest projects (in terms of KLOC, duration and commit). However, the increase of the values is still slow. The highest ratio is featured by the Commit metric. In that case, 11\% of the projects that have more than 430 commits have performed at least one migration.

These results give interesting information to answer our first research question. First of all, it is clear that few projects perform library migrations. Second, mature projects (in terms of KLOC, Commit and duration) have a higher probability to perform at least one migration. Finally, the number of commit and the duration are much more relevant to measure the maturity of a project than the KLOC for our concern. Furthermore, our results clearly show that almost 10\% of the projects that have a very long duration or a large number of commits perform migrations.

\subsection{Which libraries are migrated and how many times?}\label{sec:which}

\paragraph{Methodology.}

To verify the existence of libraries that are prone to migration, we chose to group libraries that are connected by migration rules. To that extent we computed what we call a migration graph. The nodes of this graph are libraries that have been either source or target of at least one migration. A directed arc exists between two nodes if there is at least a migration between the two nodes. To indicate the flow of migration between the different libraries, the arcs are labeled by the number of migrations that have been observed for the source and target libraries of the arcs. We then computed the connected components on the migration graph. Each connected component is a category, whose libraries of the category are the union of libraries contained in the connected component. 

\begin{figure}[h]
\centering
\includegraphics[width=0.45\linewidth]{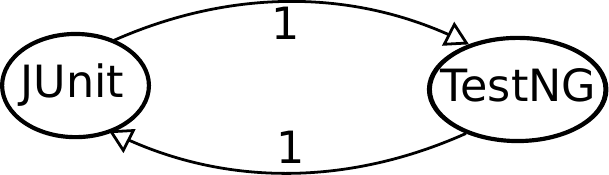}
\caption{A category formed in our example}
\label{fig:migrationgraph}
\end{figure}

A toy example of a category is shown in~\figref{migrationgraph}. This category shows that \lib{JUnit} and \lib{TestNG} are similar and that projects have performed migration in the both sides. Note that such migration graph does not mention if the same project made the two migrations or if it is two distinct projects. 

A category exhibits a set of similar libraries but also the number of migrations that have been performed among them. 

We define for a library $l \in L$ the \textit{introductions} value as the number of times any project $p \in P$ started to use $l$. We judge this concept as the most representative view of the actual usage of $l$ through several years and it allows to measure more precisely the proportion of library usage that led to a migration. Given a category, the \textit{introductions} value corresponds to the sum of the \textit{introductions} of each of its libraries.

We can then compare the number of migrations of a category with the number of library \textit{introductions} of the category. Such a ratio indicates if the category is prone to migration. 

\begin{figure}[h]
\centering
\includegraphics[width=0.55\linewidth]{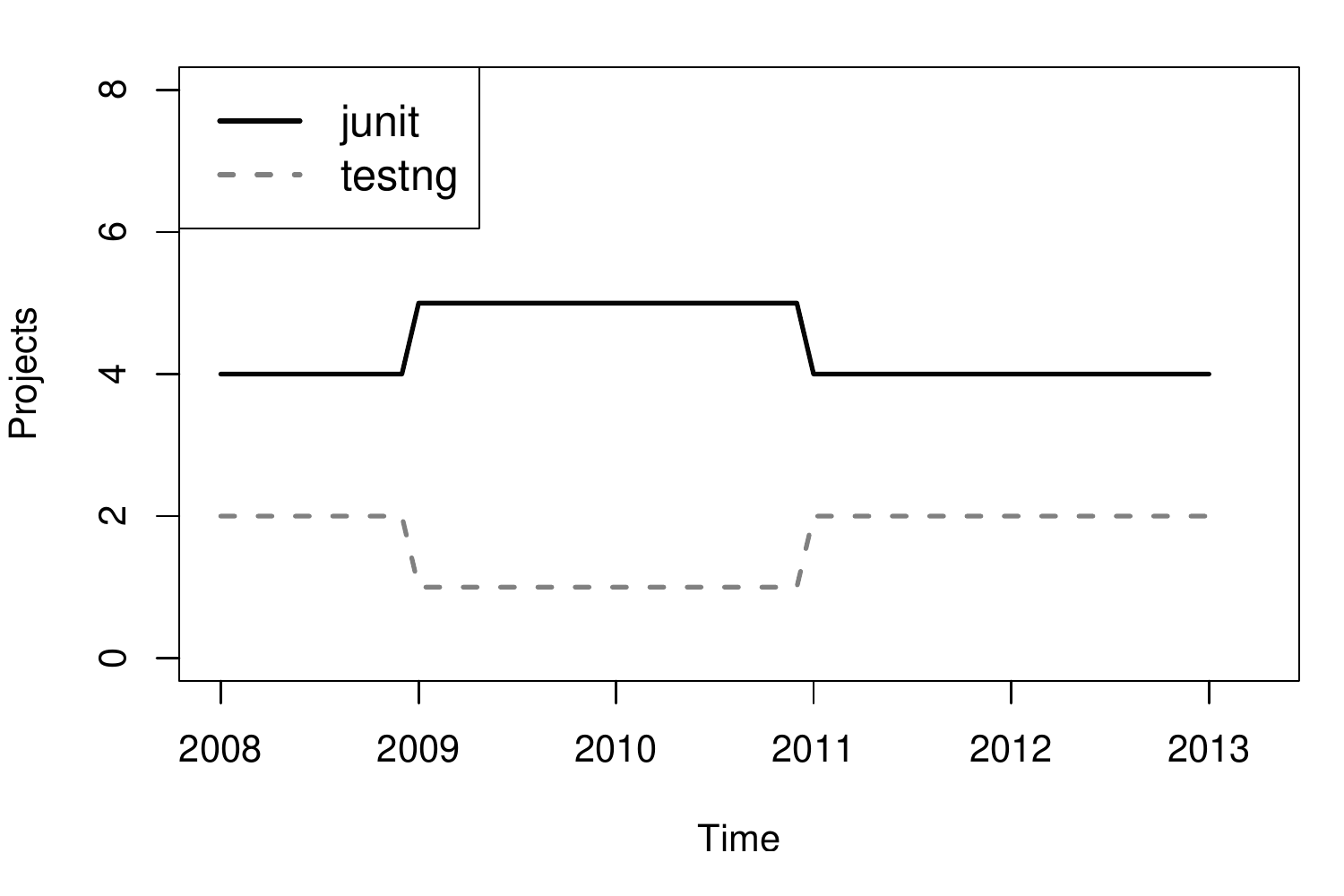}
\caption{An example of popularity-evolution diagram}
\label{fig:usage-diagram}
\end{figure}

As a complement, we introduce the popularity-evolution diagram. For a given category, it displays the evolution of the number of client projects of each library in the category. This number is computed every 2 weeks on a defined period from 2004 to 2013. An example of such diagram is exposed in \figref{usage-diagram}. A migration is characterized in this context by a loss of client for the source library and a client gain for the target library. Note that the number of \textit{introductions} in \figref{usage-diagram} is 5 for \lib{JUnit} and 3 for \lib{TestNG}, and thus 8 for the category.

\paragraph{Results.}

\begin{figure}[h]
\centering
\includegraphics[width=0.67\linewidth]{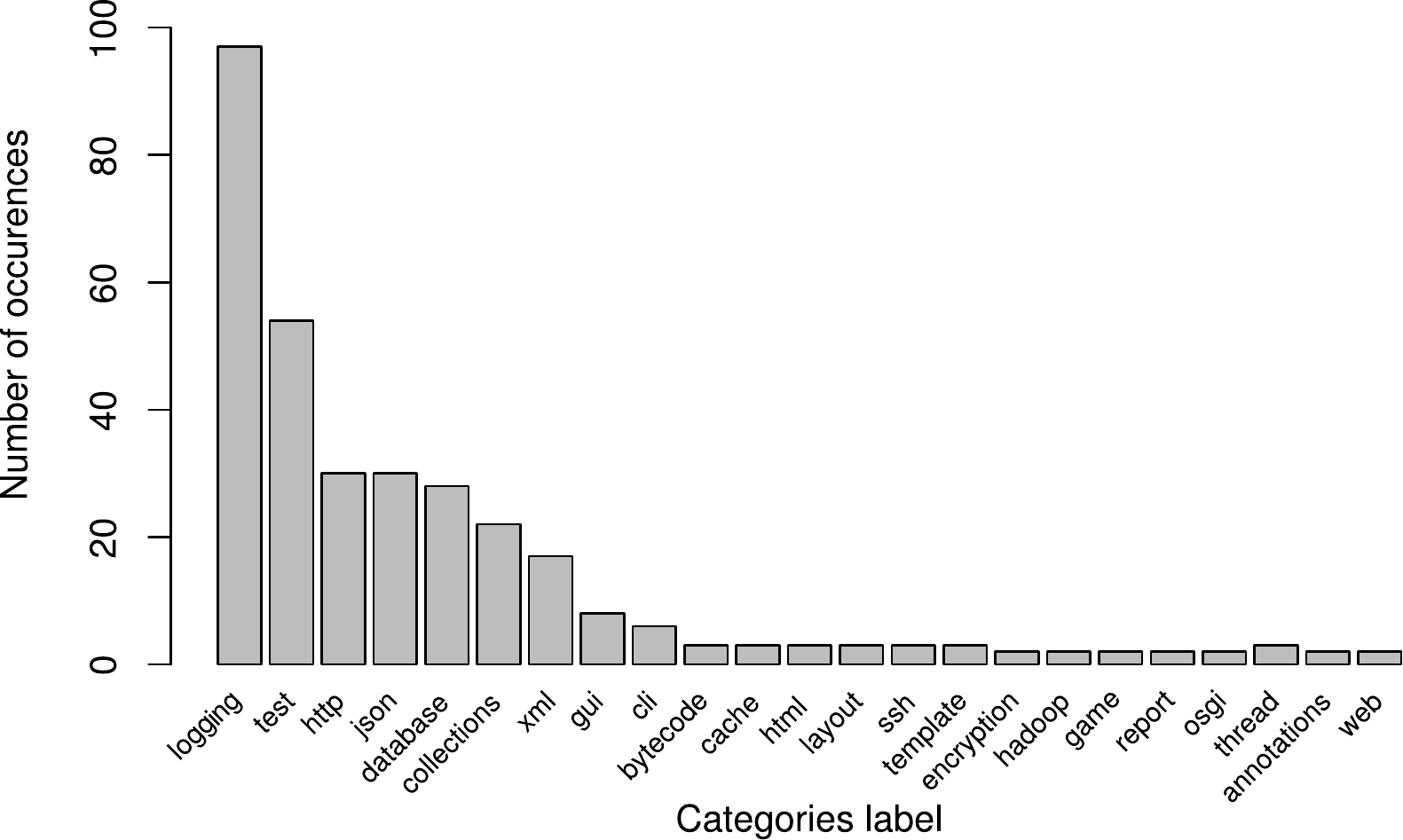}
\caption{Distribution of migrations among categories with more than 2 migrations}
\label{fig:categories}
\end{figure}

By grouping the 134 migrations rules we identified in our experiment, we obtained 38 library categories. \figref{categories} presents how migrations are distributed among these library categories. For the sake of readability, we restrain this chart to the 20 categories that have at least 2 migrations. We observe that the first 6 categories contain 74\% of the migrations, and the first 8 and 16 cover respectively 85\% and 95\% of the migrations. This result gives a first partial answer to our research question since those categories contain libraries that are prone to migration.

\begin{figure}[h]
\centering
\includegraphics[width=0.99\linewidth]{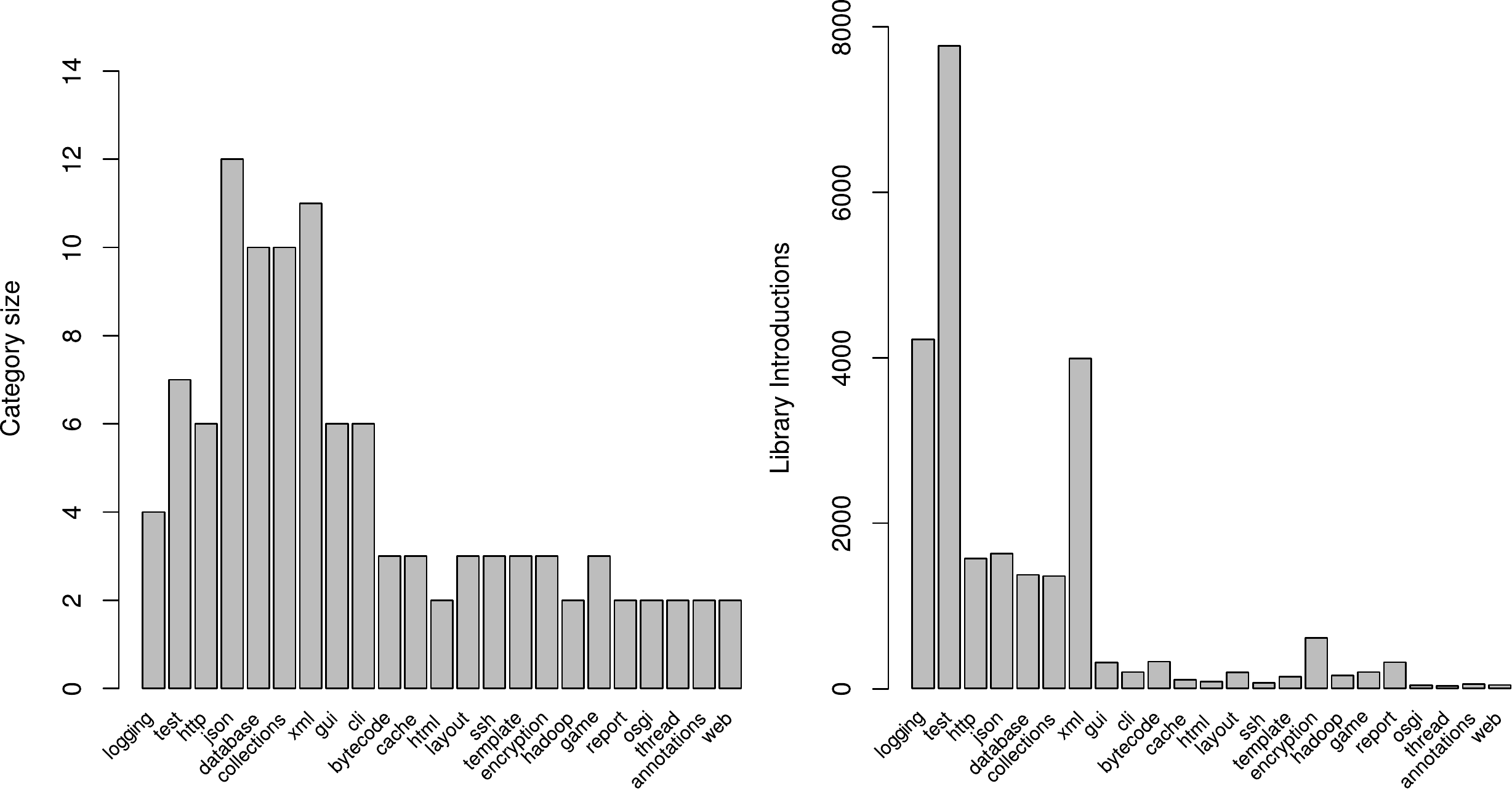}
\caption{Sizes and library introductions for the categories with more than 2 migrations}
\label{fig:categories_si}
\end{figure}

The left part of \figref{categories_si} shows the number of libraries contained within each category. This Figure clearly shows that categories contain few libraries in average (mean = 4.55). This maybe explains why few projects perform migrations as there are not so many target replacement libraries to migrate to. Further, this Figure also demonstrates  that the number of migrations per category is not correlated with the number of libraries contained in the category.

The right part of this same Figure reveals how many library introductions are counted for each category. Unsurprisingly, the two categories that contain the more migrations contain libraries that are used by most of the projects. Inversely, many projects use a \textit{xml} library but they do not perform a lot of migrations. The reason is that because the \textit{xml} category contains the native Java XML Application Programming Interface, which is used by many projects. 

These results give much information to answer our research question. There exists categories of libraries that are prone to migration. We now examine deeper the categories to better understand if there exists libraries that are prone to migration. We choose to present the \textit{logging} category as it contains most of the migration. We also choose to present the \textit{json} category as it contains few migrations but contains several libraries.

\begin{figure}[h]
\centering
\includegraphics[width=0.50\linewidth]{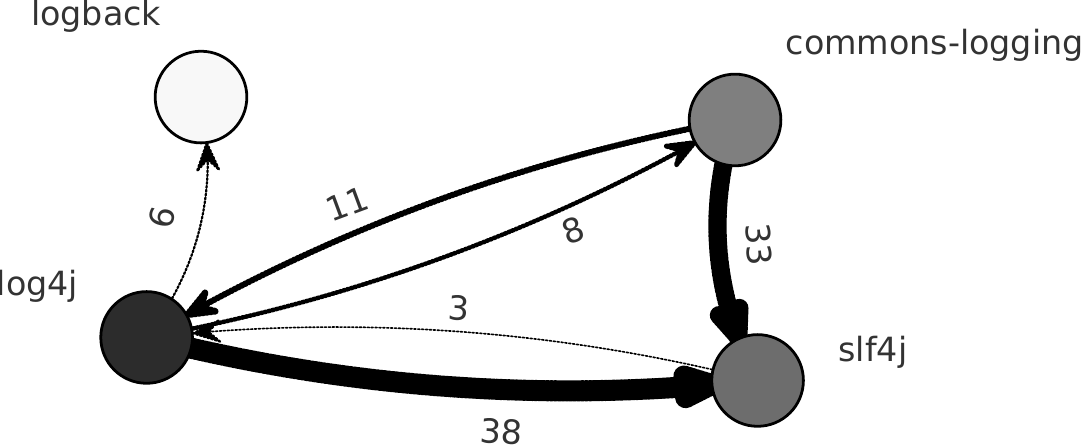}
\caption{Migration graph of the \textit{logging} category}
\label{fig:logging_graph}
\end{figure}

\figref{logging_graph} shows the migration graph that includes libraries of the \textit{logging} domain. The noticeable observation is the high number of migrations that go to the library \lib{SLF4J} (33+38=71). Moreover, many migrations go from the library \lib{Commons-logging} (33+11=44) and from the library \lib{Log4J} (38+8+6=50). We also note that \lib{LOGBack} is sometimes used to replace \lib{Log4J}, but only this library. Moreover, only 3 projects gave up \lib{SLF4J} in favor of \lib{Log4J}. 

\begin{figure}[h]
\centering
\includegraphics[width=0.60\linewidth]{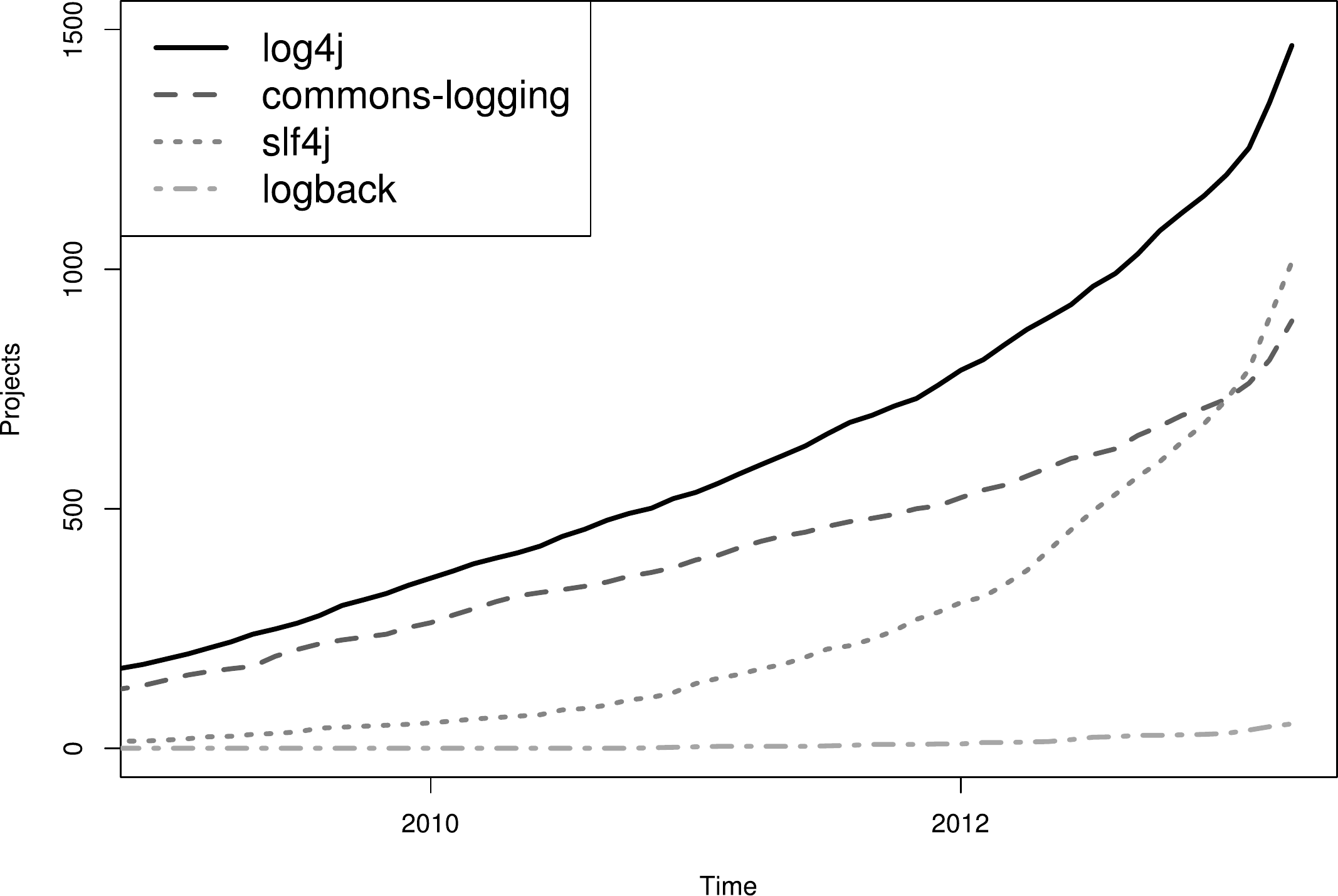}
\caption{Popularity-evolution diagram of the \textit{logging} category}
\label{fig:usage_logging}
\end{figure}

\figref{usage_logging} shows the number of projects of our corpus that use a logging library and the evolution of library usage from 2009 to 2013. This Figure first displays that \lib{Log4j}, \lib{Slf4J} and \lib{commons-logging} are the most used logging libraries in Java software development (3,425 projects in our corpus use one of these libraries). Secondly, the Figure exhibits the use increase of these three libraries, particularly for \lib{Slf4J}. Looking at \lib{Slf4J}, it appears that 1,000 projects use it currently.  Moreover, as 71 migrations go to it, this means that about 7\% ($71 \diagup 1000$) of the projects that use it were using another logging library before. \lib{Slf4J} is therefore an interesting library to consider for migration. Inversely, looking at \lib{Log4J}, even if it is currently the most popular library, there are 50 migrations that depart from it, which premises that it should be replaced.

\begin{figure}[h]
\centering
\includegraphics[width=0.70\linewidth]{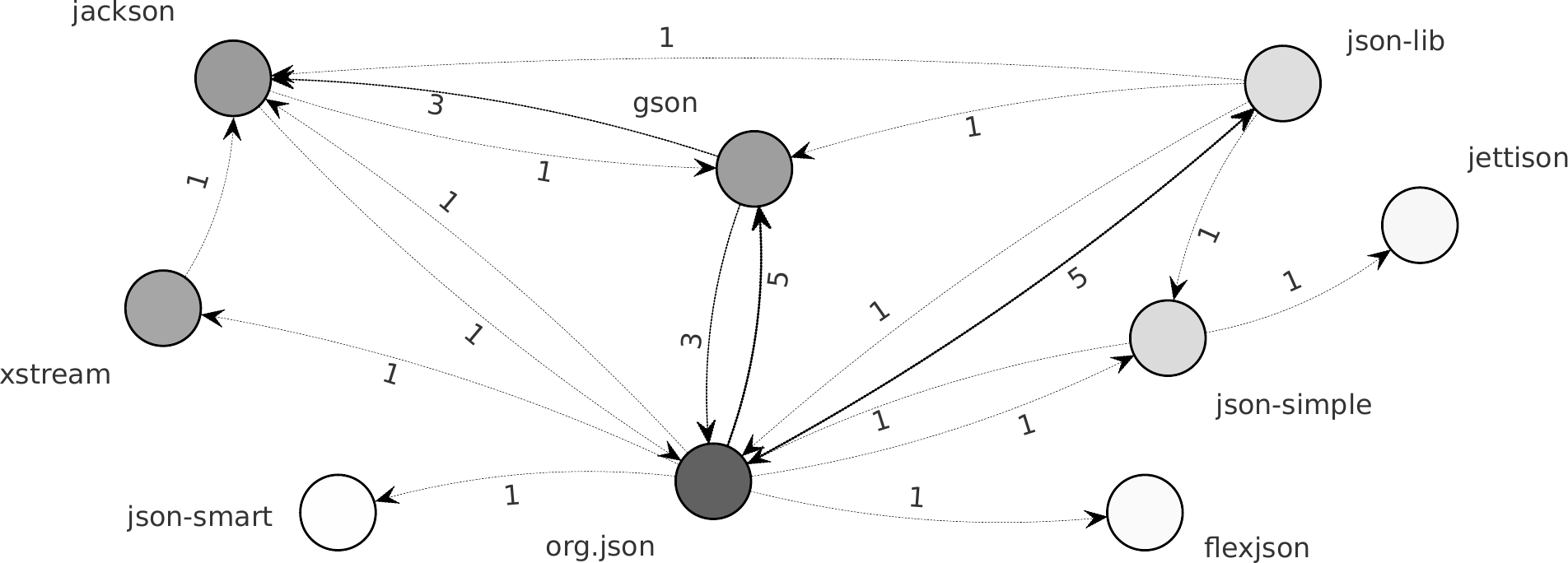}
\caption{Migration graph of the \textit{json} category}
\label{fig:graph_json}
\end{figure}

Figure \ref{fig:graph_json} shows the migration graph of the JSON category. JSON is a standard for data interchange over Web applications that is commonly and widely used nowadays. The graph contains 9 libraries and shows 30 migrations. The migrations in this graph are rather balanced. We can observe that \lib{jackson} and \lib{gson} are mainstream alternatives for library replacements. Moreover, it should be noted that \lib{org.json} is the source of 15 of migrations (50\% of the migrations).



\begin{figure}[h]
\centering
\includegraphics[width=0.75\linewidth]{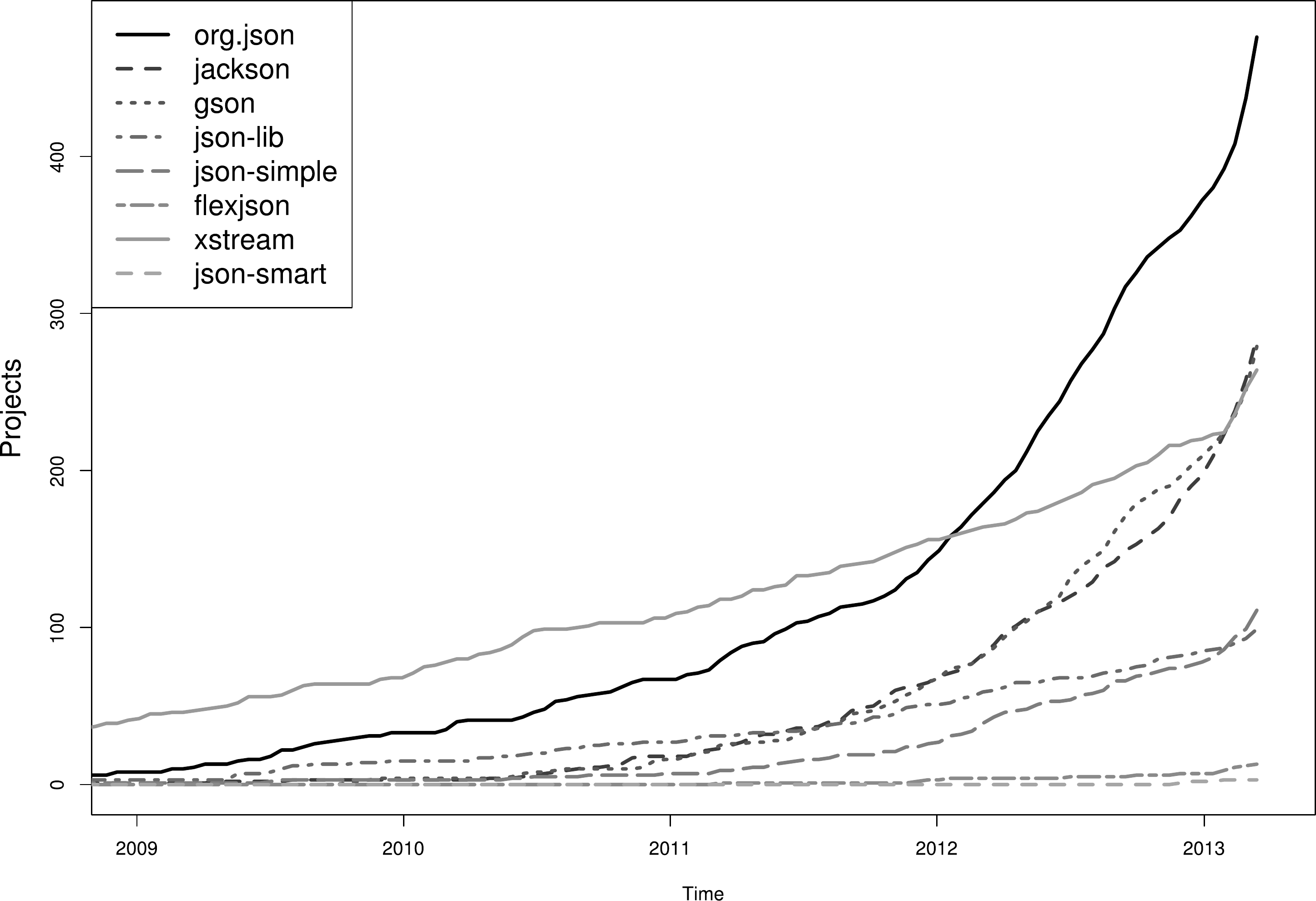}
\caption{Popularity-evolution diagram of the \textit{json} category}
\label{fig:usage_json}
\end{figure}

\figref{usage_json} shows the number of projects from our corpus that use a JSON library and the evolution of library use from 2009 to 2013. This Figure unveils that all libraries except \lib{org.json} are almost used by the same number of projects. \lib{org.json} is the most used library and has a strong increase since 2011.  However, \lib{gson} and \lib{jackson} have also a strong increase since 2012. Knowing that these two libraries are used as target to replace \lib{org.json}, they can be considered as confident candidate libraries to migrate to. 

The deep analysis of a category gives much more information to answer our research question. There clearly exists libraries that are prone to migration. This depends on the domain as well as the number of existing libraries in the domain. The \textit{logging} and the \textit{json} domains are good illustrations for that phenomena. Examining other domains highlights different libraries that are prone to migration even if few migrations are observed \footnote{\url{http://www.labri.fr/perso/cteyton/index.php?page_name=migrations}}.

\subsection{When migrations are performed?}

\paragraph{Methodology.}

To check whether the date has any influence on migrations, we compute what we call a migration-time diagram. A migration-time diagram targets one given category of libraries and presents the dates of all the migrations that happened in this category. The x-axis of the diagram presents the chronological time. The y-axis is decomposed of migration rules in the category. A point in a diagram represents a migration. This diagram allows to visualize and detect if a migration is associated to a period, and the respective trends for the source and the target as well.

\begin{figure}[h]
\centering
\includegraphics[width=0.55\linewidth]{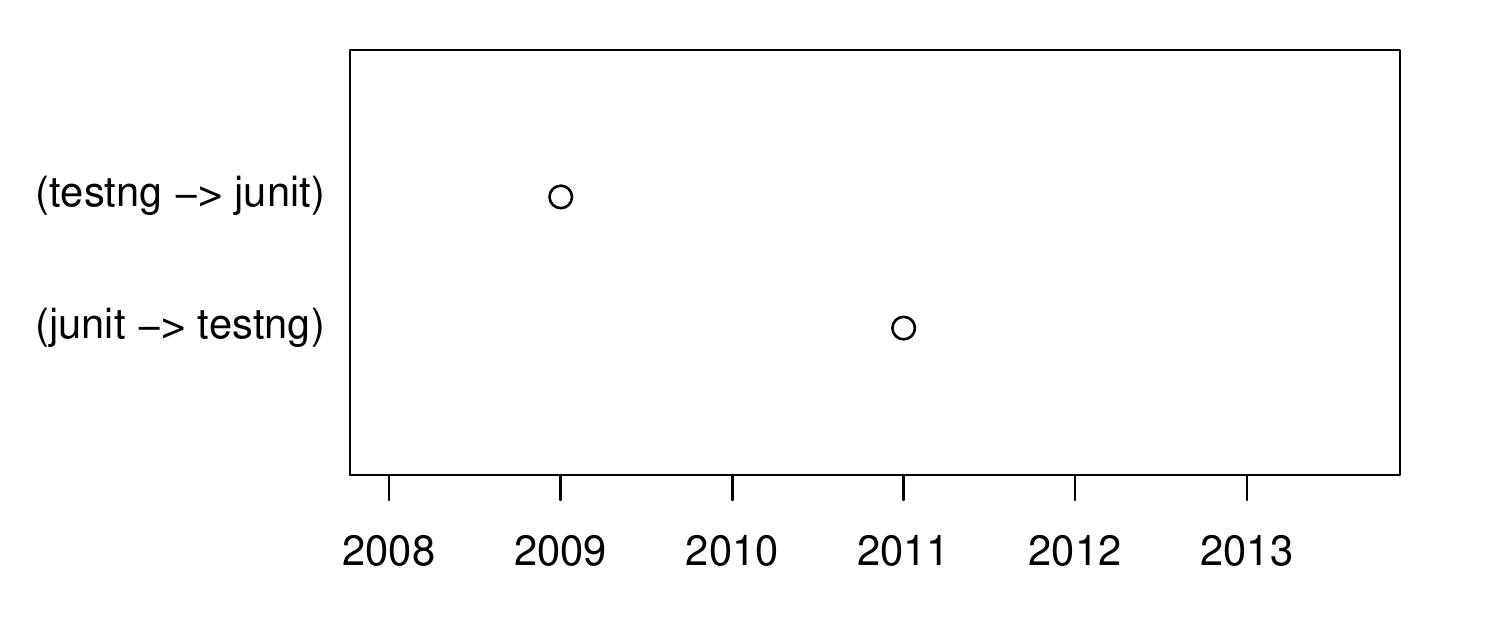}
\caption{An example of migration-time diagram for the \lib{JUnit} $\rightarrow$ \lib{TestNG} migration}
\label{fig:migration-time-diagram}
\end{figure}

Figure \ref{fig:migration-time-diagram} presents a toy migration-time diagram for the category formed in our example in \secref{which}. This diagram highlights the date when the two migrations have been performed.

\paragraph{Results.}

To check if there is any tendency regarding migrations we create a migration-time diagram for each category. For the sake of clarity, we only present in this paper the migration-time diagram of the \lib{Logging} category. The migration-time diagrams of the other categories are available online \footnote{\url{http://www.labri.fr/perso/cteyton/index.php?page_name=migrations}}.

\begin{figure}[h]
\centering
\includegraphics[width=0.90\linewidth]{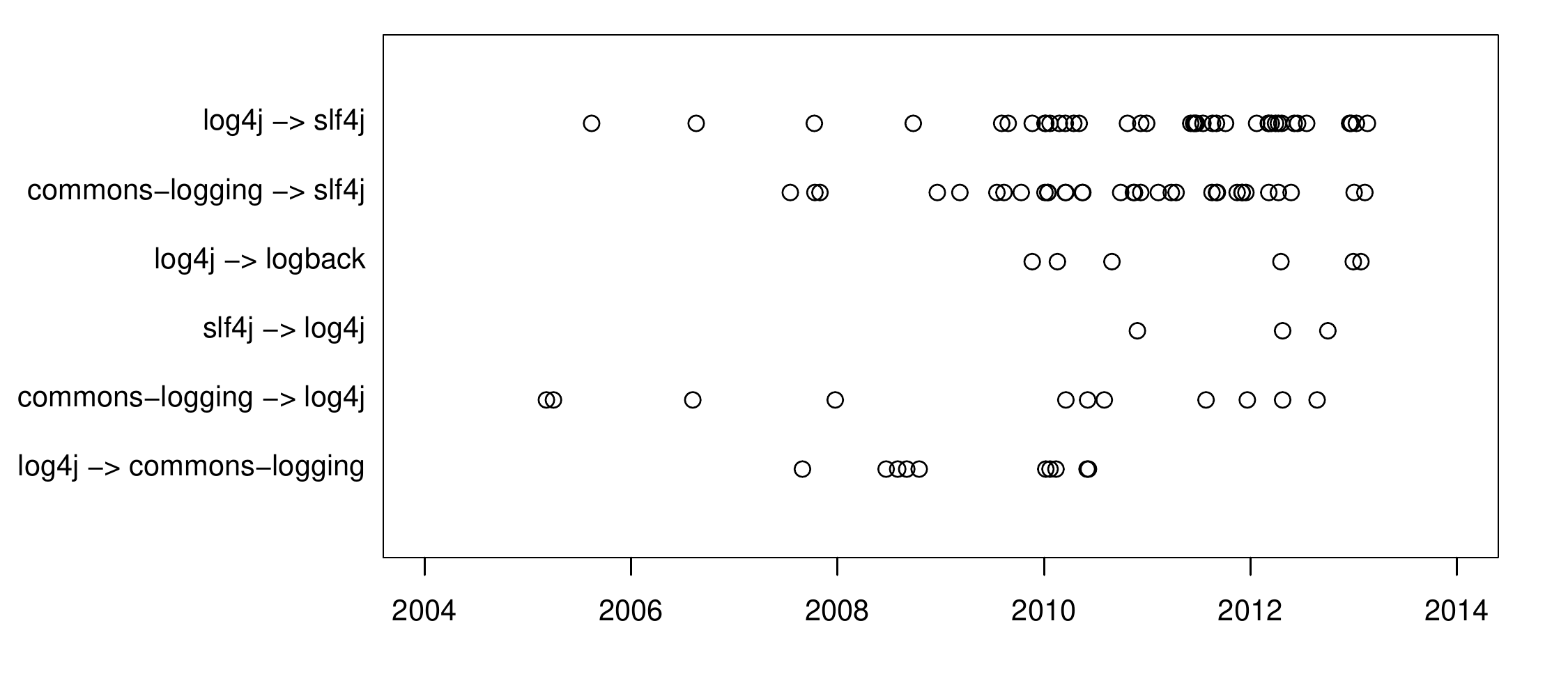}
\caption{Migration-time diagram of the \lib{Logging} category}
\label{fig:time_log}
\end{figure}

\figref{time_log} shows the migration-time diagram of the \lib{Logging} category. This category contains 6 rules that are presented on the y-axis. The diagram demonstrates that some rules are time framed such as the (\lib{log4j} $\rightarrow$ \lib{commons-logging}) one. The other rules exist during all the periods shown by the diagram. 

This diagrams also reveals how many migrations are performed for each rule. For instance, the migrations that target \lib{slf4j} have been increasingly completed since 2010. It also shows that the few migrations that target \lib{log4j} happen from time to time since 2005.

Migration-time diagrams help to answer our research question as it clearly appears that some migrations are performed during specific periods. However, these diagrams do not explain why such tendencies are observed. A further analysis of the status of libraries that were migrated as well as the status of projects that performed the migrations should be done to better understand if the period has any influence on the observed migrations..

\subsection{Why migrations are performed?}

\paragraph{Methodology.}
To identify why developers perform migrations, we decided to manually review commits logs recorded between each couple of project versions \textit{(i,j)} that contain a migration $(p,i,j,s,t)$. If a justification is provided in the log, we picked it up and tried to label it with an arbitrary category of motivation. The goal is to propose a taxonomy of the motivations found.

An example of commit log of interest found on the Web is \textit{"port logging to Slf4J (Commons-logging has classloader issues)"}\footnote{\url{http://code.google.com/p/dyuproject/source/detail?r=668}} and indicates that the library \lib{commons-logging} was dropped due to a running issue. This technique is however dependent on the quality of commit messages written by developers. 

\paragraph{Results.}

By manually reviewing the commit logs of projects that have migrated we have only found 12 logs that contain an explicit motivation. These logs are exposed in Table \ref{commits_reasons} along a motivation category. It turns out that the two most reported reasons here are \textit{Feature} and \textit{Configuration}. While the first one concerns the functionalities proposed by a library, the second tag gathers compatibility issues, conflicts due to dependencies tree and library accessibility. 

\begin{table}[h]

\centering
\footnotesize{
\begin{tabular}{lp{7cm}l}
\toprule
\multicolumn{1}{c}{\textbf{Migration}} &\multicolumn{1}{c}{\textbf{Message}} & \multicolumn{1}{c}{\textbf{Reason}} \\[2ex]
\midrule
\textit{commons-logging} $\rightarrow$ \textit{slf4j} & \textit{"replacing commons-logging with slf4j to help with osgi compliance per springsource recommendation."} & Configuration \\
\midrule
\textit{junit} $\rightarrow$ \textit{testng} & \textit{"use testng instead of junit which is a lot more configurable in test selection (and allows us to do a much better job a leaving the tree green even while developing tests that are known to fail)"} & Feature \\
\midrule
\textit{junit} $\rightarrow$ \textit{testng} & \textit{"convert tests to testng because we have groups here"} & Feature \\
\midrule
\textit{log4j} $\rightarrow$ \textit{commons-logging}& \textit{"switched from a log4j based logging api usage to commons-logging in order to allow enhanced logging techniques to be optional} & Feature \\
\midrule
\textit{log4j} $\rightarrow$ \textit{logback} & \textit{"simple migration to logback since log4j is old"} & Outdated \\
\midrule
\textit{log4j} $\rightarrow$ \textit{logback}& \textit{"changed to logback to allow change of log levels"} & Feature \\
\midrule
\textit{org.json} $\rightarrow$ \textit{gson} & \textit{"replaced json.org package with gson for license compatibility"} & License \\
\midrule
\textit{json-lib} $\rightarrow$ \textit{jackson} & \textit{"use jackson json library (reduce dependencies)"}& Configuration \\
\midrule
\textit{json-lib} $\rightarrow$ \textit{org.json}& \textit{"demonstrate bug in json-lib"} & Bug \\
\midrule
\textit{org.json} $\rightarrow$ \textit{json-smart} & \textit{"change org.json library (don't be evil license) by json smart (apache license 2.0)"} & License \\
\midrule
\textit{commons-collections} $\rightarrow$ \textit{lambdaj} & \textit{"replaced commons-collections with lambdaj since it has a more modern search syntax"} & Feature \\
\midrule
\textit{gson} $\rightarrow$ \textit{org.json} & \textit{"attempting to get into maven central repo. requires moving from gson which isn't in central (except buggy version)"} & Configuration \\
\bottomrule
\end{tabular}
}
\caption{Results of the commit logs mining to identify developers motivations to migration}
\label{commits_reasons}
\end{table}

The amount of data collected is not sufficient to propose a significant taxonomy for the motivations. Even though the logs provide rich and interesting information, we cannot infer anything based on so few data. Moreover, it is not clear why the logs do not explicitly precise the motivations of the migrations. Section \ref{sec:conclusion} discusses perspectives to better deal with this research question.

\subsection{How migrations are performed?}

\paragraph{Methodology.}
To measure efforts needed to complete a migration $(p,i,j,s,t)$, we compute how many commits were performed during the migration, how many days it took and how many developers were involved. We then compute the distribution of the migrations for each of these measures. The goal is to check whether migrations are performed within few commits, days and with few developers or if they require much more effort.

\paragraph{Results.}
\tabref{lib_mig} presents the distribution of the migrations per values for these three characteristics. It shows that 79.3\% of the migrations are achieved within a unique commit. The 20.7 \% remaining are distributed over various values, but it is still interesting to see that 6.6\% of the migrations have been completed through more than 10 commits. Further, the process of a migration is performed during one day (88 \%) or few more. Finally, a majority of the migrations is performed by a unique developer (94.2 \%).

\begin{table}[h]
\centering
-\footnotesize{
\begin{tabular}{lccccccccccc}
\toprule
 & \multicolumn{10}{c}{\textbf{Values}} \\
\cmidrule {2-11}
& 1 & 2 & 3 & 4 & 5 & 6 & 7 & 8 & 9 &  $>=$10 & Total \\
\midrule
\# Commits & 79.3 \% & 3.2 \% & 3.4 \% & 0.6 \% & 1.7 \% & 1.7 \% & 1.4 \% & 0.9 \% & 1.1 \% & 6.6 \% & 100 \%   \\
\# Days & 88.0 \% & 3.7 \% & 4.3 \% & 0.6 \% & 0.9 \% & 1.1 \% & 0.3\% & 0.3 \% & 0.0\% & 0.9 \% & 100 \% \\
\# Authors & 94.2 \% & 4.9 \% & 0.6 \% & 0.3 \% & 0 \% & 0 \% & 0 \% & 0 \% & 0 \% & 0 \% & 100 \% \\
\bottomrule
\end{tabular}
}
\caption{Distribution of library migrations per number of commits, commit days and developers  (\# : number of)}
\label{tab:lib_mig}
\end{table}


As an answer to our research question, it appears that a library migration is generally completed in one commit, in one day and thanks to one developer. In our opinion, our measures do not really reflect the real effort spent to perform the migration. This analysis confirms the assumption made in \secref{identifying_migrations:mining_lib_migrations} about libraries cohabitation existence.

\section{Limits}
\label{sec:limits}


\paragraph*{Library set}
Even though the list of libraries $L$ used to perform this study has a reasonable size, it only contains libraries that are managed by Maven. Moreover, this set does not take into account the versions of the libraries. Our approach therefore does not consider migrations across versions. Supporting such migrations would first require to identify all the versions of a library and second would require to be able to detect which version of a library a project depends on. These two issues are known to be still open~\cite{davies_software_2011}.

\paragraph*{Sampling methodology}

The corpus of projects selected for this case study has been built exclusively by querying hosting platforms. Even though Section \ref{sec:study} presents the different characteristics of the projects, we did not use any rigorous sampling approach to establish the corpus. Thus, the results of our case study cannot be generalized to any existing Java software project. 

\paragraph*{Multiple Migrations}

The approach proposed in this paper only computes migration rules of cardinality 1:1. We argue that augmented rules of cardinality n:m may exist. For instance, when a new project takes over from a no longer maintained library and split the old one into two new ones. This scenario happened in practice. Indeed, the outdated \textit{commons-httpclient} has been separated into two distinct but compatible elements, \textit{httpcore} and \textit{httpclient}. Our algorithm is in theory extensible to handle such situations, however it brings an overhead in memory space and drastically increases the number of candidate migration rules generated.

\paragraph*{Loopbacks and Bounces}

Our study does not natively consider loopbacks and bounces. A loopback is a two-steps migration observed toward the life of a project. The project first switches from a library \textit{source} to a library \textit{target}, and later in time moves back to \textit{source}. A \textit{bounce} is a migration of type \textit{x} to \textit{y}, then \textit{y} to \textit{z}, with \textit{x}, \textit{y}, \textit{z} belonging to a same category of libraries.

However, we performed a post analysis of our results to identify loopbacks and bounces. We only found two loopbacks \textit{junit} $\rightarrow$ \textit{testng} $\rightarrow$ \textit{junit} in two projects. In the first project, no information was given in the commit logs that could help us to figure out why such situation occurred. On the second project, we did find an entry in the bug tracker of the project that provides the following explanation: \textit{"Currently it is based on TestNG, however because of a number of limitations this test framework is not likely to be used by any of Hadoop's subproject. Thus, Avro will have to be start using JUnit again."}\footnote{\url{https://issues.apache.org/jira/browse/AVRO-81?page=com.atlassian.jira.plugin.system.issuetabpanels:all-tabpanel}}. The loopback is explicit but the log lacks content to point out the actual limitations mentioned by the author. One other loopback \textit{testng} $\rightarrow$ \textit{junit} $\rightarrow$ \textit{testng} was observed, but once again we could not assess why the developers made these migrations.

In the \lib{database} category, we found the following bounces : \textit{mysql-connector-java} $\rightarrow$ \textit{h2} $\rightarrow$ \textit{hsqldb},and \textit{mysql-connector-java} $\rightarrow$ \textit{postgresql} $\rightarrow$ \textit{sqlite-jdbc}. Unfortunately, these projects do not have enough material to investigate their motivations.

Therefore, loopbacks and bounces are very occasional events. In our opinion, it is not worth to investigate further this aspect of library migrations due to its rare characteristic.

\paragraph*{}

The presentation of both our approach and study is henceforth completed. \secref{rw} next presents the related work before \secref{conclusion} concludes and opens the future perspectives.

\section{Related work}\label{sec:rw}

Research has been done on software project categorization to allow searching for similar software. This problem is usually resolved by computing similarity score based on specific attributes, such as keyword identifiers as MudeBlue \cite{kawaguchi_mudablue:_????} does or API calls \cite{mcmillan_categorizing_2011}. Those techniques require either the source code or the binaries versions of a set of libraries to compute similarity scores among them. More recently, Wang et al. proposed an approach to assign tags to software using mining of existing projects tags and descriptions \cite{wang_labeled_2012}. These approaches can be used in our context to create groups of equivalent libraries but without any guarantee on the fact that a library of a group can be replaced by any other equivalent library of the group. We therefore choose to use migration graph to create categories of similar libraries. 

Mileva et al. have observed the evolution over 2 years of the dependencies of 250 Apache projects managed with the build automation tool Maven ~\cite{mileva_mining_2009}. They analyzed the maven configuration files of these projects to mine usage of API and their versions. The study shows the usage trends of different versions of several libraries. This work points out interesting cases where clients switched back to a previous version of a library they are using. We reused the idea of usage trends in the \secref{research_questions}. 

Lämmel et al. propose a large-scale study on AST-based API-usage over a large set of open-source projects~\cite{lammel_large-scale_2011}. Their work provides an insight on how a specific API is globally used by client projects. In particular, they categorize whether a client calls the API (\emph{library-like usage}) or extends it (\emph{framework-like usage}). It may be interesting to integrate such information in our library migration graph as some libraries may be more appropriate than others depending of client usage requirements.

Robillard et al. investigate the obstacles met by developers when learning an API \cite{robillard_field_2011}. Their study points out the lack of insufficient documentation or learning resources, which in our opinion can intervene in a migration context.

During a library migration, the API-level challenge is to transform the source code so that it becomes compliant with the new library. This domain aims at answering the question \textit{"How to replace a library X with Y ?"}. Bartolomei et al. have addressed this problem and studied the design of API Wrappers, which are objects that adapt and delegate the previous source code instructions towards the new API~\cite{tonelli_bartolomei_swing_2010,tonelli_bartolomei_study_2009}. The mappings are manually identified and their concern is to design such wrapper in order to obtain a compliant version of the new source code. Our approach is useful for such a problem as it identifies which libraries are source and target of migrations. It can then be used as a source of validation for the wrapper. 

The problem of updating a library has also been studied in the literature. A challenge with regard to library usage is to provide relevant snippets of code source according to the programmer's context. We distinguish two main techniques to that extent. The first one mines code that already performed an update. For instance, Schafer et al. \cite{schafer_mining_2008} examined code instantiations of two versions of a framework. This code is included with the release as test or example code. Also, SemDiff \cite{dagenais_semdiff:_2009} is a client-server connected to a framework source code repository that mines the changes and recommends modifications for a client migration. The second variety of approach requires only internal code of two API versions and applies origin analysis techniques. Hence, a graph-based representation of the code based on dependencies allows for element matching from the two versions. Some promising results have been achieved in this area \cite{wu_aura:_2010}\cite{nguyen_graph-based_2010}. Whatever the technique, our approach can be used as a massive source of data to get real library migrations and to get references of real projects that do have performed migrations. Such quantity of data could be used to validate the proposed approaches. It should be noted that a recent study from Cossette performs a retroactive study on several library changes performed manually \cite{cossette_seeking_2012}. They listed the different changes and adaptations they had to make. They argue that existing automatic approach such as the ones exposed above are not enough satisfying, since the problem of API evolution is too complex. In their opinion, this process requires at present more human intervention.

Zhong et al. proposed a Mining API Mapping approach that detects relations from two versions of an API written in different languages \cite{zhong_mining_2010}. The idea is to get client-code from the two versions and to build a transformation graph that represents the API-usage migration from one language to another.  Zheng et al. propose a cross-library recommendation tool based on Web queries~\cite{zheng_cross-library_2011}. The idea is to inquire Web search engines and to mine results proposed from the query.  One example of query could be "HashMap C\#" when looking for the equivalent for standard Java HashMap for C\#. The results are computed one by one and candidates are ranked by relevance, mainly according to their frequency of appearance. For the moment this work provides only preliminary results and queries proposed are of a coarse grain. Also, it strongly lies on Web search engines such as Google, and requires manual query writing, which can highly influence the results. Regarding our approach, this work can be used to merge equivalent libraries and then to improve library migration graphs.

\section{Conclusion}
\label{sec:conclusion}

As software intensively depends on external libraries, software developers must think about migrating libraries when they are not updated, or when competing ones appear with more features or better performance for instance. In this paper we present a study that focuses on library migration with the intent to check if they are performed on specific types of software, or for specific libraries, or during specific period, or for specific reasons, or requires specific efforts. By describing how migrations are generally managed by software projects, the objective of this study is to help software developers who are thinking about replacing the libraries of their own software.

To perform this analysis, we have defined an approach that aims at pseudo automatically identifying library migrations performed by software projects. Our approach is based on a static analysis of the source code and therefore does not depend on any tools, such as Maven, that manage library dependencies. Our approach has been prototyped for Java and successfully used on 8,600 open source software projects obtained from major hosting platforms such as GitHub, Google Code and Source Forge. Thanks to this approach, we have identified 324 library migrations, meaning that nearly 4\% of software projects performed at least one library migration. 

We then have proposed and answered research questions to better understand library migration. As a results, it clearly appears that young projects perform less migrations than older ones. Almost 10\% of the old projects perform at least one migration whereas it is less than 1\% for the young ones. Our study also shows that very few software projects perform two or more migrations during their life cycle. Further, our study shows that there are categories of libraries that are prone to migration and that some libraries within these categories are either source or target to migrations. A category gathers around 5 libraries that provide similar facilities. For instance, the \textit{logging} category gathers 4 libraries and exhibits that the  \lib{slf4j} library is currently the target of most of the migrations. 

Regarding the dates of the migrations, our study just shows that some of the migrations have been performed during specific periods. However, there is no significant data that can explain this phenomenon. Examining the reasons of the migrations, we have identified only 12 logs that give concrete explanations. Once again, this is not significant to fully answer the research question. Finally, our study shows that migrations are committed quickly (in 1 commit, one day and by one developer). This however cannot be interpreted as a measure of the effort cost for performing a migration.

The major limit of our approach is the fact that it does not support versions of library. As a consequence, an update of a library is not considered to be a migration in this study. Supporting versions is highly complex as software projects almost never define which versions of the libraries they depend on. To obtain such an information an analysis of the runtime dependencies must be done, which is still an open issue.

The results of our study should be exploitable for software developers who are looking for library recommendation. As a future work we would be interested in performing a controlled experiment with developers that want to perform a migration to check if their decision is influenced by the results of our study. For instance, we would want to check that our library migration graph can be used to identify libraries to migrate to.

We also plan to extend our approach to assist developers while they migrate their code to become compliant with a new library. As our approach identifies software projects that already performed migrations, we plan to analyze the source code of these projects before and after the migration in order to detect migration patterns. Such patterns abstract refactoring actions that must be performed to be compliant with the new library. The goal is then to automatically apply them in software projects that want to perform the same migration.

\bibliographystyle{alpha}
\newcommand{\etalchar}[1]{$^{#1}$}

\end{document}